\documentclass[a4paper,twoside]{article}

\usepackage{epsfig}
\usepackage{subcaption}
\usepackage{calc}
\usepackage{amssymb}
\usepackage{amstext}
\usepackage{amsmath}
\usepackage{amsthm}
\usepackage{multicol}
\usepackage{pslatex}
\usepackage{apalike}
\usepackage{algorithm2e}
\usepackage[bottom]{footmisc}
\usepackage{SCITEPRESS}     %

\usepackage{xcolor}

\begin{document}

\title{Multispectral Stereo-Image Fusion for 3D Hyperspectral Scene Reconstruction}

\author{\authorname{Eric L.~Wisotzky\sup{1,2}\orcidAuthor{0000-0001-5731-7058}, Jost Triller\sup{1}, Anna Hilsmann\sup{1}\orcidAuthor{0000-0002-2086-0951} and Peter Eisert\sup{1,2}\orcidAuthor{0000-0001-8378-4805}}
\affiliation{\sup{1}Computer Vision \& Graphics, Fraunhofer HHI, Einsteinufer 37, 10587 Berlin, Germany}
\affiliation{\sup{2}Department of Informatics, Humboldt University, Berlin, Germany}
\email{\{eric.wisotzky, anna.hilsmann, peter.eisert\}@hhi.fraunhofer.de}
}


\abstract{Spectral imaging enables the analysis of optical material properties that are invisible to the human eye. Different spectral capturing setups, e.g., based on filter-wheel, push-broom, line-scanning, or mosaic cameras, have been introduced in the last years to support a wide range of applications in agriculture, medicine, and industrial surveillance. However, these systems often suffer from different disadvantages, such as lack of real-time capability, limited spectral coverage or low spatial resolution.
To address these drawbacks, we present a novel approach combining two calibrated multispectral real-time capable snapshot cameras, covering different spectral ranges, into a stereo-system. Therefore, a hyperspectral data-cube can be continuously captured. The combined use of different multispectral snapshot cameras enables both 3D reconstruction and spectral analysis.
Both captured images are demosaicked avoiding spatial resolution loss. We fuse the spectral data from one camera into the other to receive a spatially and spectrally high resolution video stream. %
Experiments demonstrate the feasibility of this approach and the system is investigated with regard to its applicability for surgical assistance monitoring.}

\onecolumn \maketitle \normalsize \setcounter{footnote}{0} \vfill

\section{\uppercase{Introduction}}
\label{sec:introduction}

In recent years, multi- and hyperspectral imaging (MSI/HSI) has garnered increasing interest in the realm of scientific research, as well as in the applied fields of image analysis and monitoring. MSI and HSI are contact-free, non-invasive and non-destructive method for analyzing optical surface properties that are invisible to the human eye~\cite{LuHSIreview}. This technology is promising across a wide range of domains, including industrial~\cite{shafri2012hyperspectral}, agricultural~\cite{jung2006hyperspectral,moghadam2017plant}, medical~\cite{ClancySSI,LuHSIreview}, and security-oriented image analysis~\cite{yuen2010introduction}. While MSI captures images at specific discrete wavelengths, HSI acquires images across a vast number of continuous narrow bands and therefore attains more reliable and comprehensive information.

Hyperspectral cameras find applications across a wide spectral range from ultraviolet, through the visible range, up to infrared. The selection of the spectral range depends on the specific application with larger and narrower spectral ranges employed to obtain important information from the image data, such as biological~\cite{MuhleComparison} or chemical parameters~\cite{biancolillo2014data,borras2015data}, distinct optical characteristics of objects and their temporal variations, e.g., agricultural products~\cite{qin2008measurement}, animal organs~\cite{studier2022spectral,MuhleComparison}, human tissue~\cite{WisotzkyValidation} and industrial products~\cite{hollstein2016challenges}. 

In order to acquire hyperspectral data cubes with a large number of spectral bands (few dozens up to several hundreds), various capturing techniques exist. Line- \cite{hollstein2016challenges} or pushbroom scanning \cite{MuhleComparison} and filter-wheels~\cite{wisotzky2018intraoperative} rely on sequential scanning, and therefore need up to several seconds to acquire one complete hyperspectral data cube~\cite{ShapeyReview}. This limitation poses challenges for real-time video processing and interactive applications, where swift responsiveness is crucial. As a result, there is a growing trend towards compromising both the spectral as well as spatial resolution in order to achieve real-time acquisition of spectral data cubes \cite{MuhleComparison,EbnerSnapshot,WisotzkyTeleSTAR,wisotzky2019interactive}.

Multispectral cameras capture only specific discrete wavelengths, adopting a methodology akin to the Bayer pattern of single chip RGB cameras~\cite{bayer1976color}. These Multi Spectral Filter Arrays (MSFA) use narrow spectral masking on pixel-level on a single sensor plane, e.g., 6, 8, 16, or 25 spectral bands, also called mosaic snapshot sensors~\cite{EbnerSnapshot,WisotzkyDemosaicking}. 
The captured image data is defined by moxels (mosaic element) corresponding to the filter pattern, and stored in a data-cube representation with three dimensions, two spatial dimensions (width and height) and one spectral dimension (wavelength $\lambda$). The simplest example is similar to the Bayer pattern where one green element is replaced by another filter resulting in a $2$$\times$$2$ pattern~\cite{hershey2008multispectral}. These systems can be extended to an $n$$\times$$m$, e.g., $3$$\times$$3$, $4$$\times$$4$, $5$$\times$$5$ or even non-quadratic $2$$\times$$3$ mosaic patterns for recording wavelengths in all possible spectral ranges. The compromise made in reducing spectral and spatial resolution of multispectral cameras impairs the quality of the data. This is often addressed by spectral image enhancement algorithms aiming at spatial resolution restoration with high spectral band coverage \cite{sattar2022snapshot}. In the spatial direction, missing spectral values can be restored using interpolation or prediction techniques, also called demosaicing~\cite{arad2022ntire,gob2021multispectral}. However, this process is challenging and often inaccurate. 

In this paper, we present a multispectral stereo approach, fusing the data of two real-time capable mosaic-snapshot cameras with different spectral range into a hyperspectral data cube with high spatial and spectral coverage, circumventing the downsides of the different setups. This is achieved by adapting hyperspectral stereo vision and calibration \cite{zia20153d,Tanriverdi2019} to the specific multispectral setup in combination with multispectral demosaicing. This not only allows for high quality data fusion with high spectral and spatial resolution but also for 3D reconstruction from the multispectral recordings.

The remainder of this paper is structured as follows. The following section discusses related work in the field of MSI demosaicing and data fusion. Following, we describe our proposed spectral image fusion pipeline. Section~\ref{sec:setupdata} presents the stereo camera setup and data acquisition before section~\ref{sec:results} describes experiments and fusion results, followed by a conclusion in section~\ref{sec:conclusion}.

\section{\uppercase{Related Work}}
While multispectral imaging (MSI) records images at distinct, predefined wavelengths, hyperspectral imaging (HSI) acquires images across a broad range of continuous, narrow spectral bands. This results in HSI providing more reliable and comprehensive information. However, HSI capturing is usually based on sequential scanning, making it slow and expensive. Multispectral snapshot shot cameras employ narrow spectral filters at the pixel level on a single sensor, capturing several wavelengths at limited spatial resolution in real-time. In recent years, multispectral data enhancement or fusion approaches have gained increasing interest.

\subsection{Multispectral Demosaicing}
To enhance the spectral-spatial resolution of multispectral snapshot cameras, different concepts have been proposed in the literature, collectively categorized under the term demosaicing. In addition to the conventional interpolation techniques employed for the reconstruction of absent spectral values, such as nearest neighbor and (weighted) bilinear interpolation \cite{brauers2006color}, more sophisticated approaches have evolved incorporating domain-specific knowledge about the multispectral filter array \cite{Mihoubi2017}. 

In recent years, the application of deep neural networks (NN) has emerged in the field of hyperspectral imaging. These networks have been leveraged to predict HSI data from MSI or even classic RGB image data~\cite{Arad2022Challenge,Arad_2020_CVPR_Workshops}. Methods based on deep neural networks have been shown to achieve superior accuracy than classical approaches, especially in the delineation of spatial and spectral boundaries. However, there exist central challenges associated with these approaches. The main challenge pertains to the lack of training data and the substantial reliance of the method. %
This implies that the spectral characteristics of individual scenes are learned for interpolation purposes and thus establishing a critical dependence between the planned application and quality and quantity of the training data. As a result, suboptimal outcomes and inaccuracies in hyperspectral data may manifest if the training data is inadequately chosen, of poor quality or insufficient in scope \cite{wang2021deep}.
Further, achieveing real-time capability as well as mitigating artifacts like blurring remain unsolved challenges in the integration of deep neural networks into hyperspectral imaging processes \cite{lapray2014multispectral}.

\subsection{Multispectral Data Fusion}
Data fusion has been used to combine data from an MSI, an RGB, or a grayscale sensor with an HSI camera~\cite{guo2022multispectral,oiknine2018dictionary,Tanriverdi2019} in order to enhance the spatial resolution of the HSI data. The combination of different types of optical sensors with different spectral and spatial resolutions, especially for remote sensing, leads to the problem that these sensors provide data in different modalities depending on their sensor characteristics. However, the integration of data of different sources can lead to an improved accuracy in data analysis and interpretation \cite{maage2008regression,Tanriverdi2019}. 
Most approaches employ sophisticated setups using a beam splitter or a shifting unit to ensure that both cameras are virtually aligned on the exact same optical path \cite{guo2022multispectral}. However, the use of a beam splitter carries the risk of lower intensity and interpolation of the spectral information leading to blurring artifacts~\cite{arad2016sparse}. Other approaches use light-field imaging \cite{manakov2013reconfigurable,wisotzky20233d,maccormac2023lightfield}.
In order to interpret and analyze the data from different sensors in a cohesive manner, methods for data fusion are needed \cite{forshed2007evaluation,biancolillo2014data}. While general fusion techniques have undergone significant development and find applications in various fields ranging from satellite-based Earth observation to computer vision, medical imaging, and security, the fusion of data from multiple sources with different spatial, spectral, and temporal resolutions remains challenging \cite{zhang2010multi}.

In recent years, research efforts have been directed towards the extraction of 3D image data from multispectral images \cite{sattar2022snapshot}. This involves the calculation of the 3D shape of the object's surface by analyzing multispectral images captured from different camera views \cite{zia20153d,klein2014stereo}. It is worth noting that all existing methods in this domain employ the same multispectral imaging range for both viewpoints or use overlapping spectral ranges to allow 3D analysis in this range \cite{sattar2022snapshot,bradbury2013multi,shrestha2011one,genser2020camera}.

The process of data fusion can be grouped into three strategies: low-level, mid-level and high-level \cite{pohl1998review}. The low-level data fusion operates at pixel level, straightforwardly concatenating the raw data from all sources pixel by pixel into a single matrix. This process can be executed without additional preprocessing; although occasional data adjustments may be necessary to ensure compatibility \cite{borras2015data}. In contrast, mid-level data fusion, also referred to as feature level fusion, involves the extraction of relevant features from each data source individually. These features are depending on the inherent characteristics of the data, e.g., spectral intensities, edges or texture. Subsequently, these features are fused into a single dataset for further processing, such as classification or regression \cite{dong2009advances}. At high-level fusion, processing is performed on each data block independently and the results are combined to achieve the final decision. Therefore, it is often denoted as decision level fusion \cite{zhang2010multi}.

\section{\uppercase{Hyperspectral Fusion}}

In order to achieve a hyperspectral image with high spatial resolution and real-time capabilities, we combine two multispectral snapshot mosaic cameras with $n$ and $m$ spectral bands, covering different spectral intervals into a stereo setup. To fuse the captured data of both cameras into one HSI data cube, we first demosaic the snapshot mosaic images (section \ref{sec:demosaic}), before calculating a dense registration between the two stereo views (section \ref{sec:featuredetection}) and fusing the different spectral data based on the calculated warps.

\subsection{Demosaicing}
\label{sec:demosaic}
We employ demosaicing as an essential processing step, serving two key purposes: enhancing the visual quality of the images into full spatial resolution and facilitating accurate feature detection. This step is particularly crucial due to the substantial intensity differences between neighboring pixels caused by the different filter responses and specific spectral behaviors, i.e., due to the structure of an MSFA. Attempting feature detection and registration on the single-channel mosaic pattern would therefore be impractical. In addition, demosaicing enables us to accommodate sensors with different mosaic sizes in the stereo setup, e.g., as in this work $n=4$$\times$$4$ and $m=5$$\times$$5$.

\begin{figure}[ht]
\centering
	\includegraphics[width=\columnwidth]{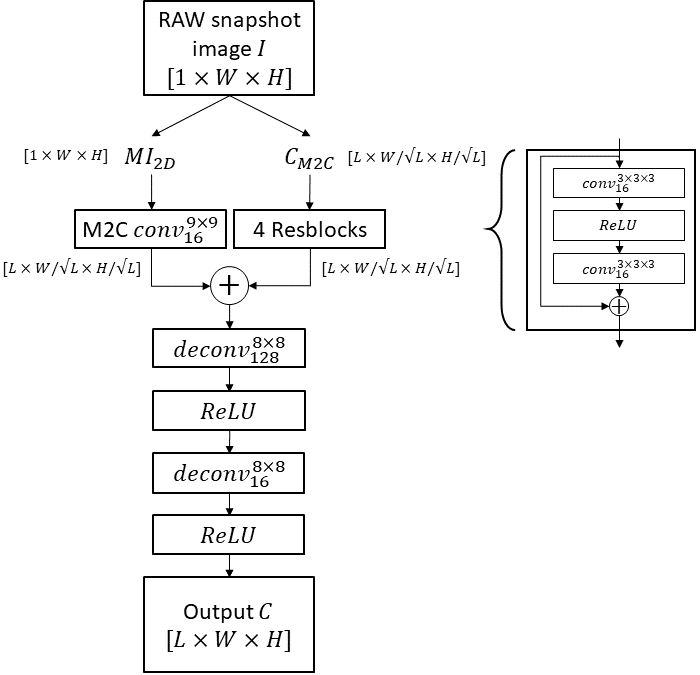}
	\caption{The demosaicing network. Two parallel feature extracting layers using a mosaic-to-cube converter (M2C) on one side and ResNet blocks on the other side are used followed by a feature adding and two deconvolution (\textit{deconv}) layers to upsample the spatial dimensions of the image.}
	\label{img:demosaicking}
\end{figure}

In our demosaicing process we opt for a CNN architecture with parallel building blocks following the methodology detailed in~\cite{WisotzkyDemosaicking}.
The network uses a 3D CNN for spectral refinement as well as a convolutional layer for mosaic-to-cube interpolation, followed by 2D deconvolutional layers for spatial upsampling at each spectral band, see Fig.~\ref{img:demosaicking}.

\begin{figure*}[ht]
\centering
	\includegraphics[width=.9\textwidth]{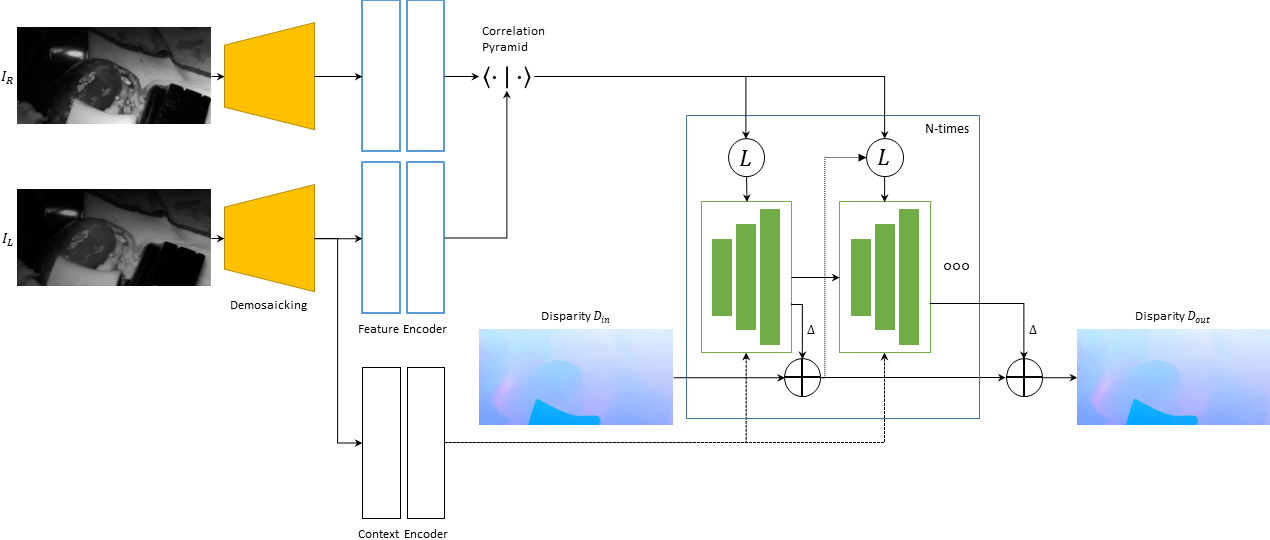}
	\caption{The entire pipeline to calculate disparity. The first part includes parallel demosaicing of both camera images. Then a feature extraction follows using a mixture of convolutional layers and ResNet units. As last step the disparity is calculated iteratively by using several convolutional blocks.}
	\label{img:combNetwork}
\end{figure*}

\subsection{Disparity Calculation} \label{sec:featuredetection}
Other work to calculate the image disparity has focused on two key parts: (1) feature matching and (2) image regularization \cite{genser2020camera,shrestha2011one,sattar2022snapshot,Wisotzky2023Curr}. 
Given two images, feature matching aims to compute a matching cost between a pair of image patches. In general, this is an optimization problem maximizing the visual similarity with given constraints on the 3D geometry. Optical flow formulates the same problem based on the entire image content combining both parts. %
Therefore, we use a adapted approach which iteratively computes the optical flow between two images. 

In order to calculate the disparity between the two stereo images, we use a pixel-wise alignment based on optical flow calculation. However, optical flow is based on the brightness constancy assumption. Our two multispectral images ($I_L$ and $I_R$) hold different numbers of channels as well as different information from different wavelengths. For each demosaiced view, we therefore first average the spectral channels after channel-wise normalization, resulting in single channel images $I'_L$ and $I'_R$. Due to different spectral coverage of the two cameras, these images hold different reflectance information and thus violate the brightness constancy assumption of optical flow. However, our experiments showed that the networks we used can adapt to this. We selected three different models from the literature as basis for the optical flow model: RAFT \cite{teed2020raft}, DIP \cite{zheng2022dip} and MS RAFT+ \cite{jahedi2022high}. Each of these models has been adapted to allow gray-scale inputs. The feature detection is performed on each image individually using a Feature Encoder, see Figure \ref{img:combNetwork}. Subsequently, an iterative process determines the disparity information. We use the disparity of the previous image pair for an initialization of this process, reducing the number of iterations of the disparity calculation to two.

Evaluation of these networks has performed using widely used benchmark datasets: FlyingThings3D \cite{mayer2016large}, MPI Sintel \cite{Butler:ECCV:2012} and KITTI optical flow 2015 \cite{Menze2015CVPR}.

\subsection{Data Fusion}
Utilizing the dense disparity information, the spectral information is fused  by transferring the spectral information from the left view ($I_L$) to the right view ($I_R$).
To achieve optimal aligned spectral data, the fusion process involves consideration of various factors that influence the spectral response in the two diverse cameras. These factors involve spectral filter responses of the individual bands, camera parameters and illumination conditions. To account for these factors, we apply spectral correction and reconstruction as elaborated in \cite{WisotzkyValidation,MuhleComparison}. 
Finally, the spectral bands are sorted according to their specific wavelength.

This results in a hyperspectral data cube with high spatial as well as spectral resolution ($n+m$ spectral bands). In addition, the disparity map provides depth information, such that accurate 3D reconstruction can be derived from a calibrated stereo system.

\subsection{Stereo Calibration}
In order to retrieve reliable spatial information from the disparity map, a calibration process of the stereoscopic system is essential \cite{rosenthal2017kalibrierung}.
For this purpose, we use a checkerboard pattern \cite{eisert2002model}. 
The automated calibration principle we used in this work has six main steps and needs only to be performed once the stereo-system has been changed.
\begin{enumerate}
	\item \textbf{Capturing.} Acquisition of multiple ($n$) calibration views ($I_C$) guaranteeing that images stay in focus resulting in a calibration set $I_{Cj}$ where  $j = [1,n] := \{j \in \mathbb{N} |1 \leq j \leq n\}$.
	\item \textbf{Feature detection.} Detection of 2D feature points $F$ in $I_{Cj}$ analogue to Sec.~\ref{sec:featuredetection} and assign a 2D-to-3D mapping between captured views $I_C$ and reference model views $I_R$ resulting in a unique ID-based feature point set $N(I_{Cj}, I_{Rj}) = {F1(id, x, y), F2(id, x, y), \ldots} \rightarrow {F1(id, x, y), F2(id, x, y), \ldots}$.
	\item \textbf{Intermediate calibration and pre-orientation.} Using $N(I_{Cj}, I_{Rj})$ to compute an approximated camera pose $P(I_{Rj})$ via a least-square optimization describing a rigid body transformation $[R|t]_{j}$ where $R$ is a $3\times 3$ rotation matrix and $t$ a $1\times 3$ translation vector, between the captured view $I_{Cj}$ and the model reference view $I_{Rj}$. Then
	\begin{equation}
	P(I_{Rj}) = \arg\min( N(I_{Cj}, I_{Rj}) )
	\label{eq:recoCorr}
	\end{equation}
	minimizes the L2-norm between detected feature points and reprojected 2D points of reference model features using the current $[R|t]$ estimations.
	\item \textbf{Single camera calibration.} We estimate the intrinsic camera parameter by a linear gradient-based estimation of camera pose and lens parameters using the pre-orientated poses of step 3. The process is performed independently for the left and right camera. Currently, we estimate the focal length $f_x, f_y$ and three radial distortion parameters $k_1, k_2, k_3$ and ignore the principal point and tangential lens distortion.
	\item \textbf{Stereo camera calibration.} The estimated camera view pose from step 4 allows a straightforward computation of the camera extrinsic parameter, which describe relative transformation from right camera $C_R$ to the left camera $C_L$ to complete the stereo calibration. 
	\item \textbf{Calibration validation.} We validate our calibration routine with two geometrical statistics in 3D space.\\ (a) Measurement accuracy: the distances between feature points are known with high accuracy. The distances between a preselected set of distances is computed for the reconstructed points and compared to known ground-truth values.\\ (b) Planarity constraint: The distance of the reconstructed point to the reconstructed object plane is an indicator of planarity of the reconstructed object, which needs to be fulfilled, as the original calibration target is also planar. 
\end{enumerate}

\section{\uppercase{Stereo Setup and Experiments}}
\label{sec:setupdata}

To select the best performing optical flow model, we performed different evaluations on the three adapted models. For consistency, we used the data from Flying Things3D and MPI Sintel for training and evaluated on MPI Sintel and KITTI optical flow 2015 data. All images are converted from RGB to gray-scale after channel-wise normalization. To simulate the differences of the captured reflectance intensity in the two cameras, we applied different modulations on the three channels of one camera image during the gray-scale conversion. One of the following functions is randomly selected and applied for each channel: identity $y=x$, square root $y=\sqrt{x}$ and $y=1-\sqrt{1-x}$, exponentiation $y=x^n$ and $y=1-(1-x)^n$ with $n=2$ or $n=4$, and logarithm $y=\log_2(x+1)$ and $y=1-\log_2(2-x)$.

\begin{figure}[t]
\centering
	\includegraphics[width=.9\columnwidth]{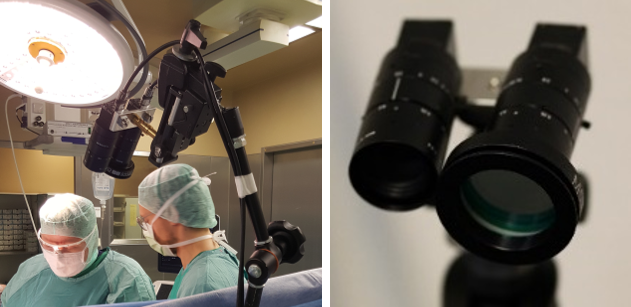}
	\caption{The stereo-HSI-setup. Left: the setup mounted on a surgical retractor system in the operation room. Right: detailed view of the stereo-cameras mounted on the rigid trail.}
	\label{img:setup}
\end{figure}

\begin{table*}[h]
\caption{The adapted and analyzed optical flow models for disparity calculation. The models are trained either only on FlyingThings3D or on a mixture of FlyingThings3D and Sintel. The analysis in terms of end-point-error (EPE) is done on Sintel and KITTI data.}
\centering
\begin{tabular}{lrcccc}
Model    & Training       & EPE Sintel orig & EPE Sintel proc & EPE KITTI orig & EPE KITTI proc \\ \hline \hline
RAFT	 & FlyingThings3D & 7.87       & 8.05    & 5.65    & 5.55\\
DIP      & FlyingThings3D & 9.34       & 10.90   & 8.84    & 9.74\\ \hline
RAFT     & mixed          & 3.01       & 3.66    & 1.67    & 2.83\\
DIP      & mixed          & 4.67       & 4.49    & 7.52    & 8.34\\
MS RAFT+ & mixed          & 5.36       & 6.01    & 2.11    & 4.21\\
\end{tabular}
\label{tab:NetAna}
\end{table*}

The best performing model is then integrated in our pipeline to evaluate the functionality and intraoperative usability of our multispectral image fusion. We captured intraoperative stereo data with two different multispectral snapshot cameras (XIMEA GmbH, Germany) mounted on a rigid trail, see Fig.~\ref{img:setup}. The left camera is a $4$$\times$$4$-VIS camera covering the spectral range from $\lambda = 436$~nm to $\lambda = 650$~nm with $16$ spectral bands and a sensor resolution of $1024$$\times$$2048$~pixels. The right camera is a $5$$\times$$5$-NIR camera covering the spectral range from $\lambda = 675$~nm to $\lambda = 975$~nm with $25$ spectral bands on a $1080$$\times$$2045$~pixels sensor. Both sensors have a pixel size of $5.5$$\times$$5.5$ $\mu$m and the cameras hold a F$2.8/75$~mm optics (Ricoh, Japan) \cite{MuhleComparison}. The $5.5$$\times$$5.5$ camera includes a band-pass filter on top of the optics.

The distance between both cameras is $a = 6$ cm. The cameras are rotated around the image plane for about $10$ degree. This allows to capture the same scene using a working distant (WD) of about $d = 25-35$ cm. Both cameras are frame-wise synchronized.

To provide the full spectral information from $\lambda = 400-1000$~nm for each point in the scene, the captured spectral data of both cameras have to be fused into one consistent hyperspectral data cube. To allow a correct processing on the fused spectral data, we apply a color correction to the raw data as desribed in .%
A white reference image is acquired directly before or after the actual capturing to correct the present illumination setting and achieve the reflectance information. This is of high importance, especially for LED illumination, since a large difference in intensity exists between the individual recorded spectral bands (within on camera as well as between both cameras).

After the demosaicing step, both images are cropped to identical sizes of $1020$$\times$$2000$~pixels for better image handling. Due to performance reasons, we warped the $16$ bands of the $4$$\times$$4$-camera and concatenated the $16$ spectral points for all pixels having disparity information at the beginning of the $25$ bands of the $5$$\times$$5$-camera to achieve $41$ spectral bands per pixel.

To demonstrate the feasibility of our approach, we conducted two types of experiments. In the first one, we acquired $n=25$ images of the checkerboard from different viewing angles and WD to perform the spectral-stereoscopic calibration. Further, images of a color chart (ColorChecker Classic, X-Rite Inc.) have been acquired to allow comparison of the fused spectra to ground truth information.
For the second experiment, we acquired medical data to show the spectral data fusion working in sensitive applications as image-guided surgery. We captured stereo-spectral images during two different types of surgeries: parotidectomy (ENT surgery), cf.~Fig.~\ref{img:warping} and kidney transplantation, cf.~Fig.~\ref{img:final}. Both types of surgery are performed as open surgeries. In these cases, the surgeon usually does not use an optical device as microscope or endoscope. Only magnifying glasses may be used at times to enlarge anatomical structures.

For illumination, a halogen light source is used for the calibration experiments. During surgery, the present surgical light is used, which always comprehensively covers the interesting parts of the scene with high brightness.

\section{Results}
\label{sec:results}

\subsection{Network Selection}
The evaluation of the three adapted networks are summarized in Tab.~\ref{tab:NetAna} and showed the best performing results for our adapted RAFT. In most of the cases, the dataset without gray-scale modulation (orig) during RGB to gray-scale conversion shows better results in comparison to the data with modulation (proc). However, for all training modalities (either only on FlyingThings3D or a mixture of FlyingThings3D and MPI Sintel) RAFT performed best regardless on which data we evaluated (MPI Sintel or KITTI). Therefore, we selected RAFT for further usage.

\subsection{Calibration}

\begin{figure}[ht]
\centering
	\includegraphics[width=.9\columnwidth]{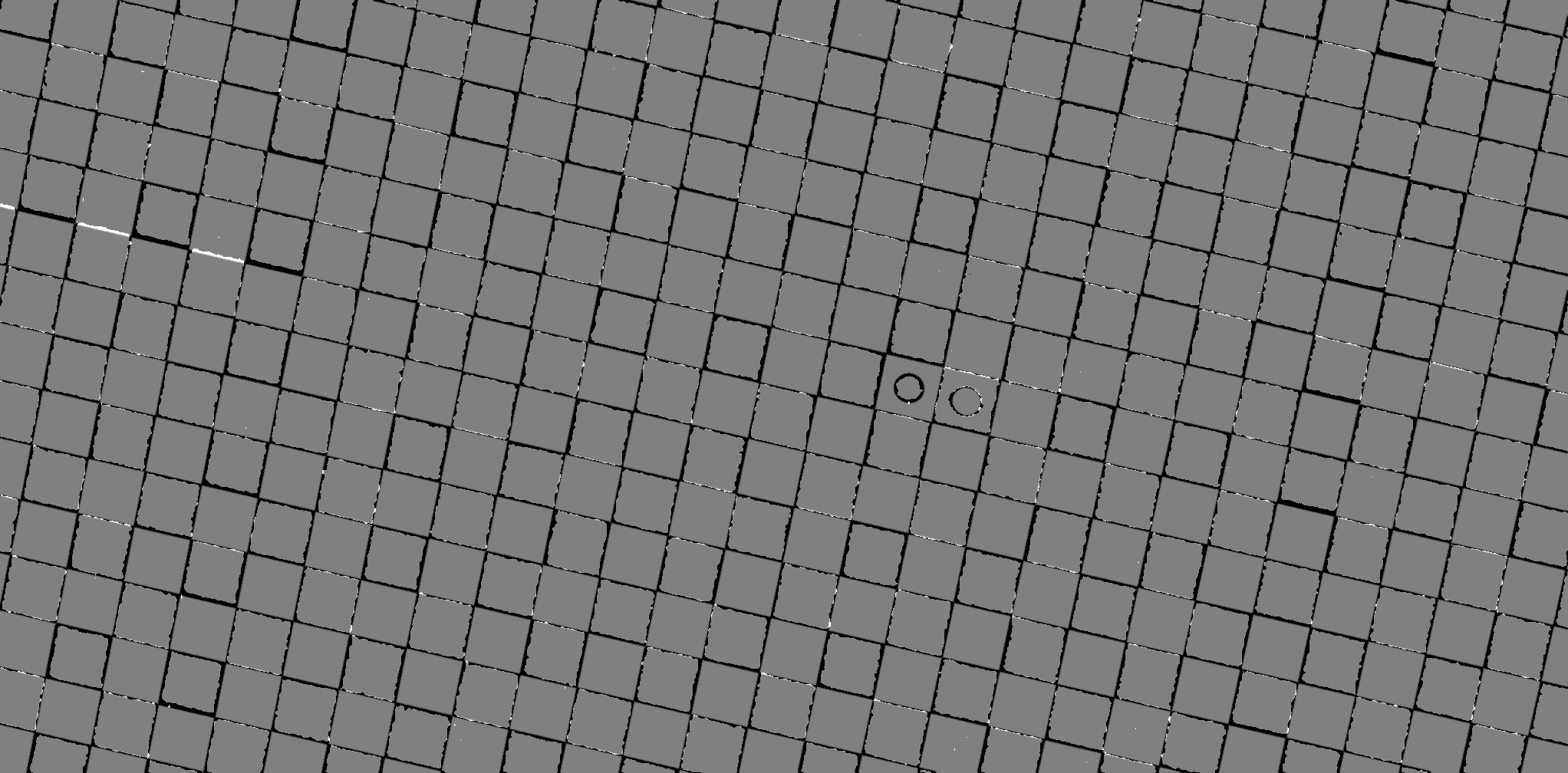}\\
    \includegraphics[width=.9\columnwidth]{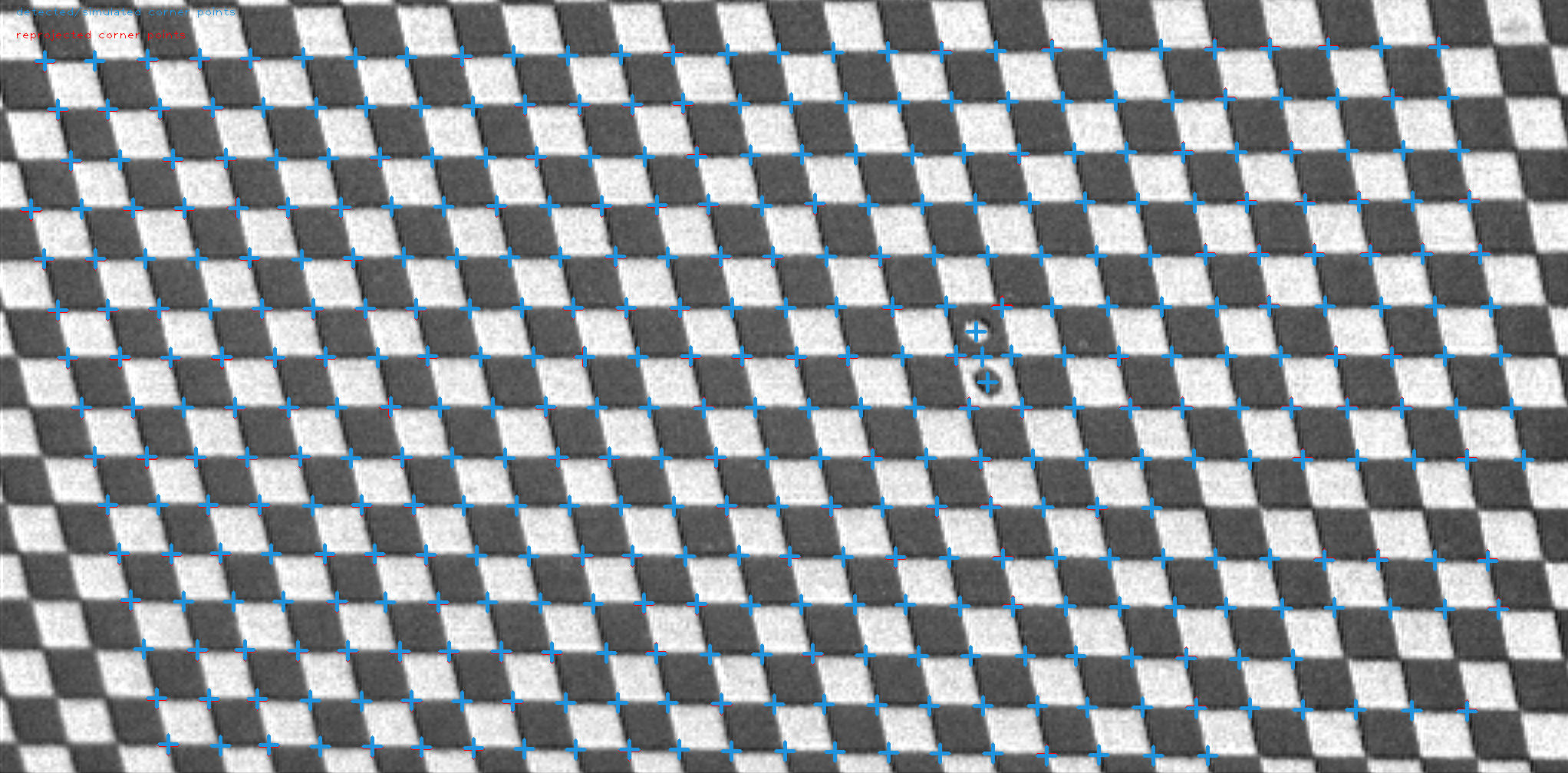}
	\caption{Calibration results. The top figure shows an example of a difference image between the projected ground truth checkerboard data and a captured left ($4$$\times$$4$ mosaic snapshot) camera view. The bottom figure shows an example of the re-projected corner points of the checkerboard. The detected corner points (blue) fit almost perfectly with the re-projected corner points (red).}
	\label{img:reproject}
\end{figure}

We performed the described calibration process ten times on the acquired calibration images. In comparison to the measured camera to camera distance of $a = 6$ cm, we achieved $a_{calib} = 5.987$ cm when we averaged all spectral channels of each camera during the performed calibrations. For individual channel calibration, we achieved an distance up to $a = 7.29$ cm. In addition, the convergence angle $\theta$ between the cameras is determined as $\theta = 3.5\deg$ in the calibration with all channels averaged. For individual channel calibration, we determined $\theta$ in the ranges of $\theta = [6.96, 11.24]\deg$. As mean stereo re-projection error $1.50$ px was achieved, cf.~Fig.~\ref{img:reproject}.

For the focal length $f$, we calculated $\bar{f}$ by averaging all spectral channels of each camera resulting in $\bar{f}_{left} = 79.69$ mm and $\bar{f}_{right} = 78.52$ mm. But for narrow-band multi- and hyperspectral imaging, focal length generally exhibits a distinct wavelength dependence called chromatic aberration. Therefore, we analyzed the focal length for each individual spectral band. %
In that case, the focal lengths ranged between $f_{left} = 81.23$ mm to $88.60$ mm and $f_{right} = 71.13$ mm to $78.08$ mm.

\subsection{Hyperspectral Fusion}
The captured image data of the color chart and seven surgeries in two different surgical fields have been analyzed. It was possible to reconstruct a dense disparity map and fuse the spectral data for all cases. To evaluate the quality of the fused spectral data, we analyze the captured color chart images.
For the case of a plane color chart the disparity information are not of large interest but the spectral reconstruction of the single color tiles. %
These reconstructions were previously analyzed in detail and could be reproduced~\cite{MuhleComparison,WisotzkyValidation}.

\begin{figure}[t]
\centering
	\includegraphics[width=\columnwidth]{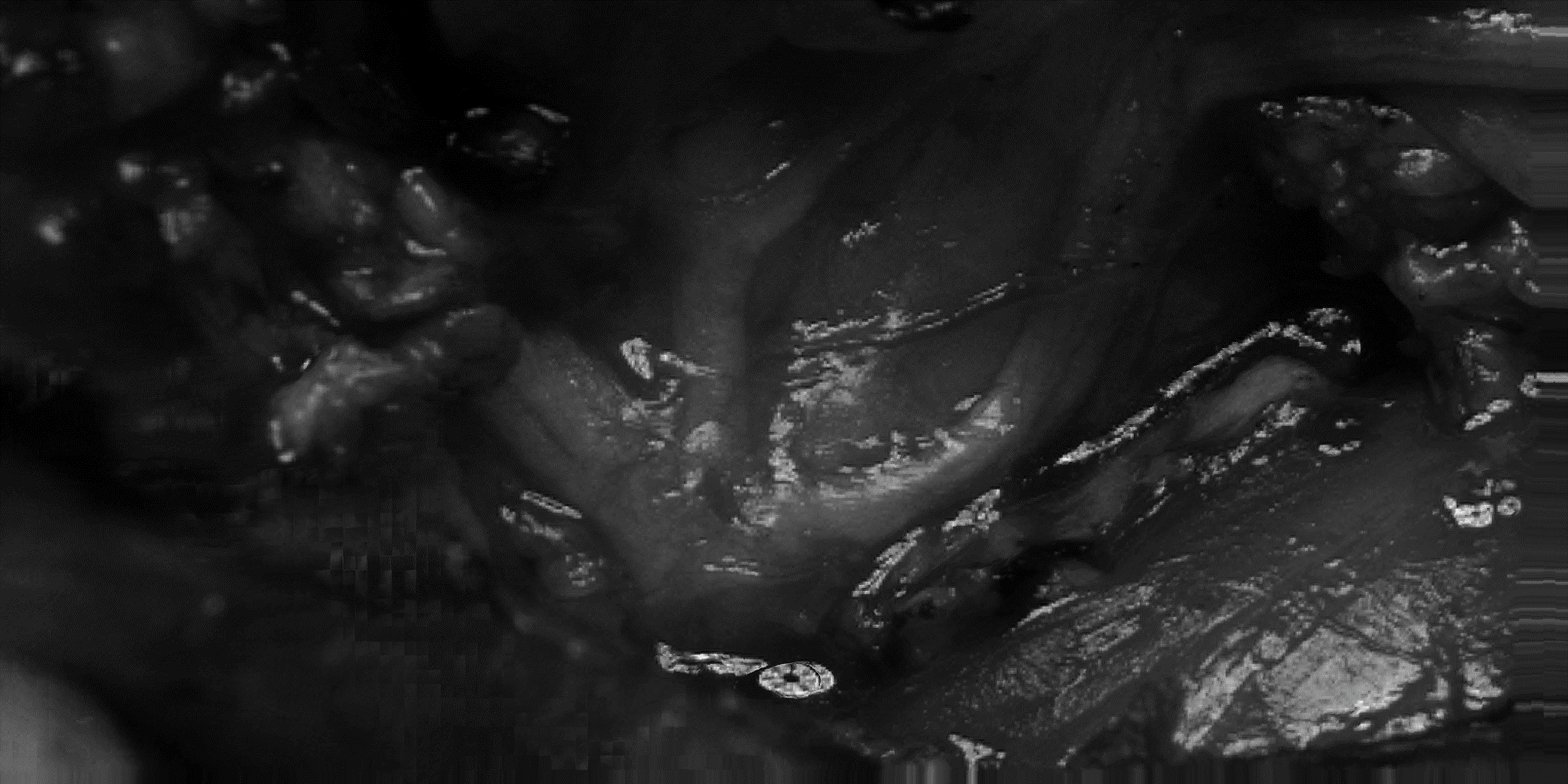}
	\caption{One spectral channel of the right $4$$\times$$4$ camera view after warping into the other camera view. Due to the warping, some areas at the image borders hold no data and are repeated by neighboring pixels. To avoid this effect during analysis, the final data cube is cropped.}
	\label{img:warping}
\end{figure}

\begin{figure}[h]
\centering
	\includegraphics[width=\columnwidth]{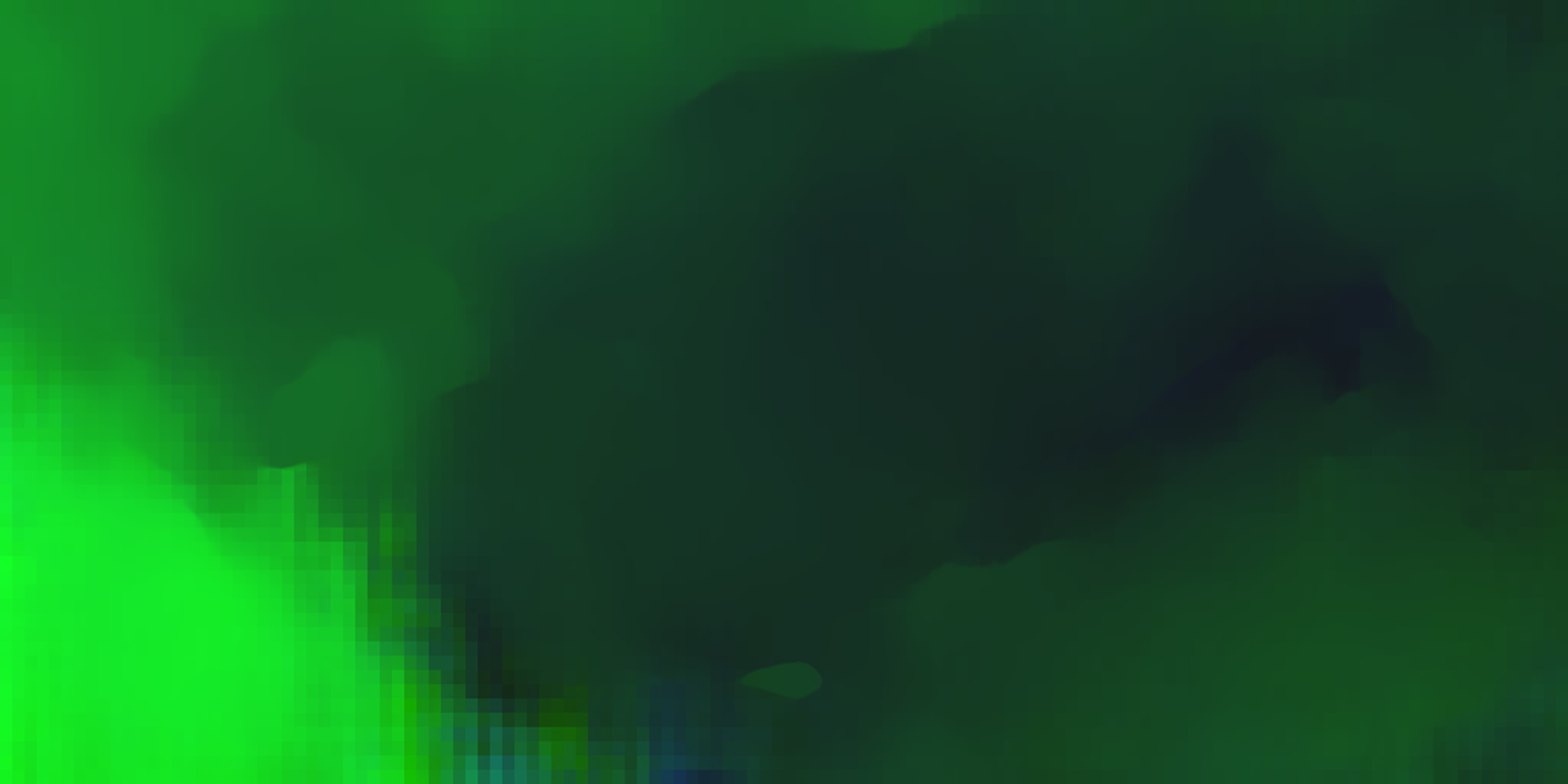}
	\caption{The disparity map of the case shown in Fig.~\ref{img:warping}. Dark pixels refer to regions far away from the camera and bright pixels refer to regions closer to the camera. The white area on the left holds no information due to the stereoscopic baseline shift. In addition, due to the baseline shift and missing information, there is also an incorrect prediction of a region distant from the camera system (red pixels) that is actually close to the camera system.}
	\label{img:disparity}
\end{figure}

The final hyperspectral data cube is always smaller than the multispectral-stereo input images, since the baseline of the stereo system leads to a small scene shift as well as to occlusions in the two multispectral-snapshot images. Therefore, the rectified and warped image show empty image borders, cf.~repeating pixel lines in Fig.~\ref{img:warping} at the bottom and right borders. The dense disparity map of this case is presented in Fig.~\ref{img:disparity}. Artifacts are visible in the specular highlights. In these cases, the correct distance from the scene to the camera cannot be determined. However, this is not of high relevance, since these regions do not contain any usable spectral information and are not usable for further analysis. 

Using the calculated disparity map, the captured spectral data can be taken pixel-wise from the $4$$\times$$4$-camera and fused in front of the $5$$\times$$5$-camera data. Thus, for each pixel having a valid disparity correspondence a full reflectance spectrum from the captured scene can be calculated. Using this information, a realistic RGB representation can be calculated and the object spectra in the full range of $\lambda = 400$ to $1000$ nm can be used for analysis.

\begin{figure}[t]
\centering
	\includegraphics[width=.9\columnwidth]{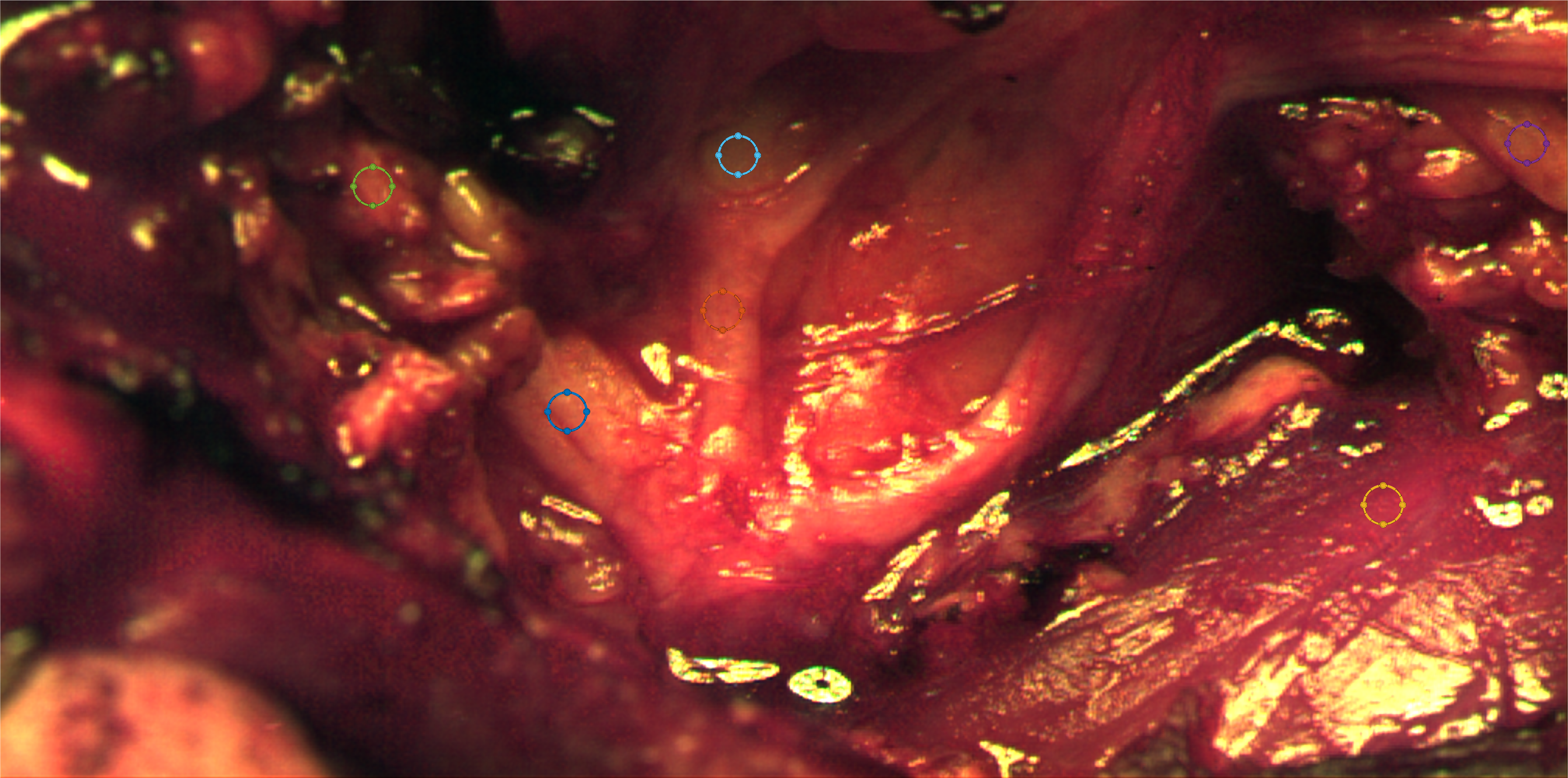}
	\includegraphics[width=.9\columnwidth]{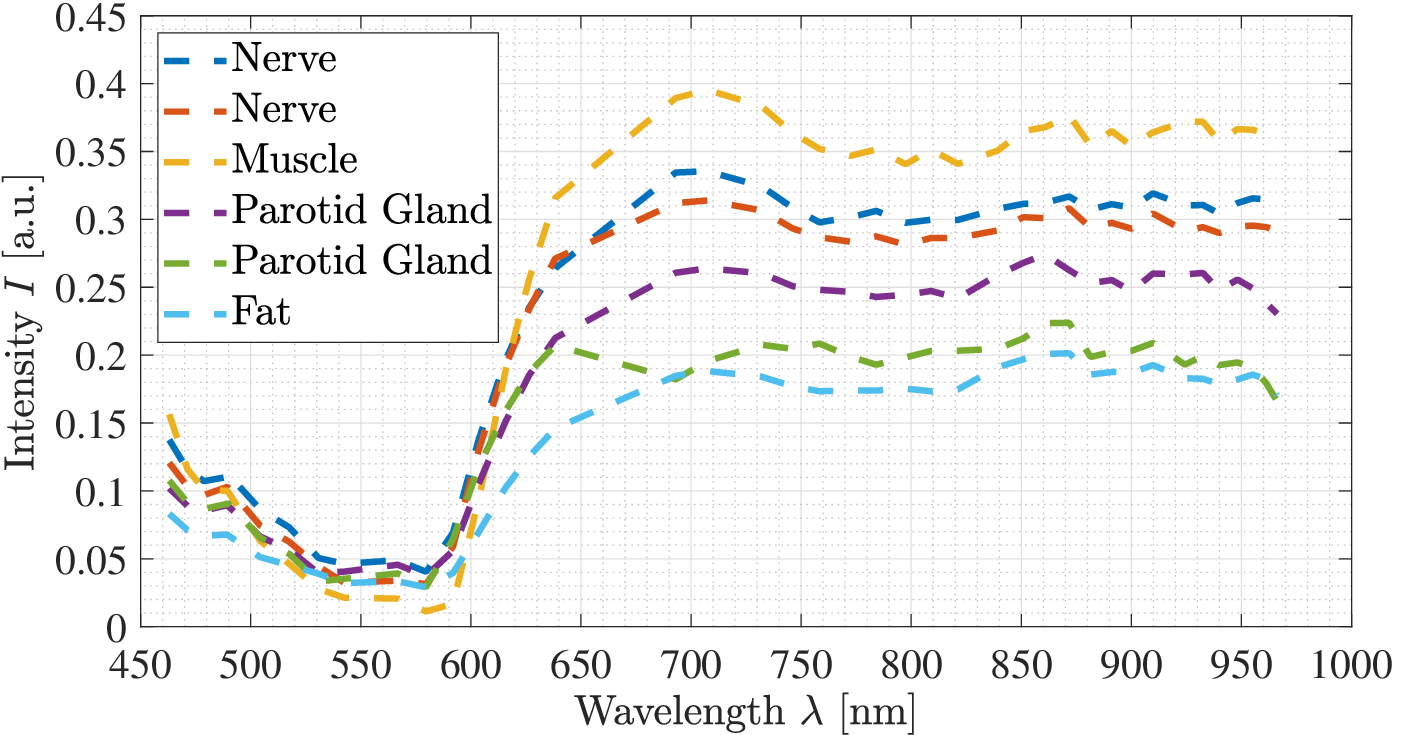}
	\caption{Due to the fusion of the spectral information into one camera view, a realistic RGB representation can be calculated (top view). Further for all regions with disparity information, the spectral information in the VIS-NIR range can be extracted, cf.~colored circular ROIs in the top view and spectral plots in the bottom view.}
	\label{img:spectralplot}
\end{figure}

\begin{figure*}[h]
    \centering
	\includegraphics[width=.9\columnwidth]{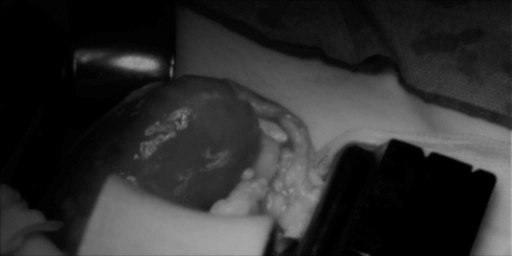}
	\includegraphics[width=.9\columnwidth]{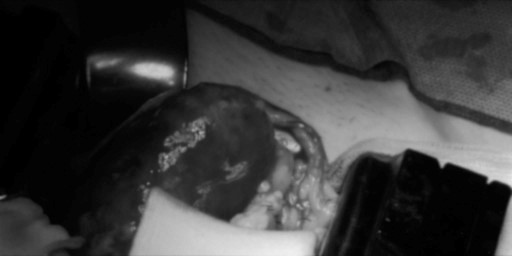}
	\includegraphics[width=.9\columnwidth]{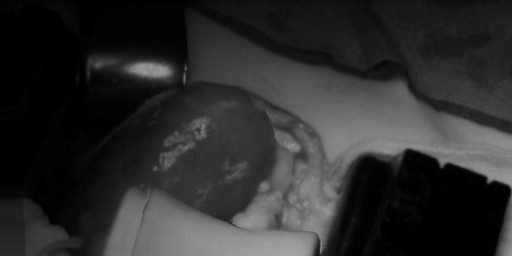}
	\includegraphics[width=.9\columnwidth]{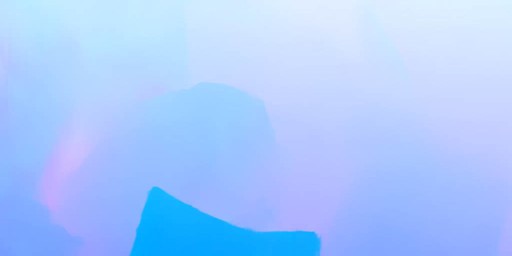}
	\caption{The final data fusion of data captured during kidney transplantation. Top left: one spectral channel of left camera view, demosaicked. Top right: one spectral channel of right camera view, demosaicked. Bottom left: the top left view (left camera) warped into right camera view. Bottom right: Flow field/disparity map of scene, blue color corresponds to small camera-object distance and red is larger camera-object distance.}
	\label{img:final}
\end{figure*}

The resulted reflectance spectra of one surgical scenes (parotidectomy) are presented in Fig.~\ref{img:spectralplot}. In this example, six regions showing four different tissue types are selected and the according spectra plotted. Identical tissue types, regardless of the image region and camera-surface distance at which they were located, result in a similar spectral reflectance spectra, while the spectra of different tissue types show clear differences in their reflectance behavior. All reconstructed spectral tissue curves are in accordance to reported spectral tissue behavior \cite{EbnerSnapshot,MuhleComparison,studier2022spectral,bashkatov2005optical}.%

In Fig.~\ref{img:final} all results of another surgical use-case (kidney transplantation) are presented. The top left shows a spectral band of the demosaicked left view. The calculated depth map (top right in Fig.~\ref{img:final}) is dense and matches the visual appearance of the scene (red pixel are close to the camera and blue pixel are far away). Only at the left edge and next to the liver hook (white region at the bottom center of the scene) no depth information can be extracted due to the stereoscopic occlusions. For a part of the liver hook, the depth information could not be reconstructed correctly due to the occurring of total reflection in one of the stereoscopic views. However, the overall data fusion and spectral reconstruction works well as shown for two different spectral bands in the bottom row of Fig.~\ref{img:final}.

\section{\uppercase{Conclusions}}
\label{sec:conclusion}

In this work, we present an image/spectral fusion method for stereoscopic multispectral snapshot images covering entirly different spectral ranges. This allows a continuous capturing and processing of intraoperative images to achieve a hyperspectral data cube. 
The hyperspectral data cube shows high spatial resolution with about $2000$$\times$$1000$ pixels and a good spectral resolution with $41$ spectral bands in the interval of $\lambda=400-1000$ nm. This spectral resolution with the present sampling is sufficient to reconstruct a spectrum in the VIS-NIR region for robust analyses, which is comparable to spectra extracted from other hyperspectral cameras \cite{MuhleComparison}. Furthermore, our system combines advantages of different HSI and MSI systems from the literature without showing their disadvantages \cite{EbnerSnapshot,ClancyFW,ClancySSI}. 
A problem that quickly arises is when the objects to be analyzed are not visible in both images due to their spectral properties. Then it is difficult to register both images and fuse the spectra. This would inevitably lead to worse results, but could certainly be solved with complex and controlled illumination solutions.
However, this problem is not likely to occur in the application area of image-enhanced surgery addressed here.

In addition, the system allows continuous capturing and reconstruction of the 3D surface texture, e.g., the situs in intraoperative imaging, which thus creates a significant value increase. This makes it possible to continuously monitor the present objects and their spectral behavior over time. 
For example, for image-guided surgery anatomical structures and physiological changes can be analyzed at the same time during the entire procedure, if necessary. Further, it is possible to measure the object sizes or register these precisely to other image data, e.g., preoperative CT. Thus for the clinical application, the surgeon can be comprehensively provided with objective data. Other comparable systems only allowed to reconstruct few spectral bands and/or 3D surface, and were not real-time capable \cite{LuHSIreview,WisotzkySensor}.
Our system can continuously capture and process data with low latency showing the potential of real-time analysis as the pipeline is only partially implemented for GPU usage.

\section*{\uppercase{Acknowledgments}}
This work was funded by the German BMBF under Grant No.~16SV8602 (KIPos) and by the German BMWK under Grant No.~01MT20001D (Gemimeg II). The authors state no conflict of interest. Informed consent has been obtained from all individuals included in this study. The research related to human use complies with all the relevant national regulations, institutional policies and was performed in accordance with the tenets of the Helsinki Declaration, and has been approved by the Charit\'e ethics board.

\bibliographystyle{apalike}
{\small
\bibliography{literature}}

\end{document}